\documentclass
[prb,twocolumn,notitlepage,showpacs,a4paper,nobalancelastpage]{revtex4-1}%
\usepackage{amsfonts}
\usepackage{subfigure}
\usepackage{amsmath}
\usepackage{amssymb}
\usepackage{graphicx,epstopdf}
\usepackage{appendix}
\usepackage{array}
\usepackage{color}
\usepackage[normalem]{ulem}
\usepackage{CJK}
\providecommand{\U}[1]{\protect\rule{.1in}{.1in}}

\newcommand\rmv{\bgroup\markoverwith {\textcolor{red}{\rule[0.5ex]{2pt}{0.4pt}}}\ULon}

\newcommand{\fudan}{\affiliation{Department of Physics and State Key Laboratory of Surface Physics, Fudan University, Shanghai 200433, China}}

\def\strutdepth{\dp\strutbox}
\newcommand\marginhack{\strut\vadjust{\kern-\strutdepth\hacksize}}
\newcommand\hacksize{\vtop to \strutdepth{ \baselineskip\strutdepth \vss\llap{{\tiny {\margintext}\quad }}\null}}
\newcommand\margintext{margin texts big and wide and wide }

\newcommand{\abs}[1]{\left| #1 \right|}

\newcommand{\bz}{\ensuremath{\mathbf z}}

\begin{document}

\begin{CJK*}{UTF8}{gbsn} 
\title{Spin snake states with spin-orbit and Zeeman interactions in an inhomogeneous magnetic field}
\author{Vahram Grigoryan}
\email[Corresponding author:~]{vgrigoryan@fudan.edu.cn}
\fudan

\begin{abstract}
We study the spin edge states, induced by the combined effect of Bychkov-Rashba spin-orbit, Zeeman interactions and inhomogeneous magnetic field, exposed perpendicularly to two-dimensional electron systems. We calculate analytically the spectrum of the spin edge states (the spin snakes orbits) in systems where the magnetic field exhibits a discontinuous jump in the transverse direction and changes its sign at the magnetic interface. The obtained magnetic spin edge states in a 2DES exhibit several interesting properties: in particular, electrons reverse their propagation direction at finite values of their momentum.
\end{abstract}

\pacs{72.25.Dc, 72.10.-d, 73.63.Hs, 73.21.Fg}
\maketitle
\end{CJK*}
Investigations in semiconductor structures of reduced dimensionality are frequently connected with the use of a homogeneous magnetic field, \cite{1} which, in addition to the lateral confinement, quantizes the carrier motion also in the plane normal to the magnetic field. In hybrid ferromagnetic-semiconductor nanostructures (HFSN) \cite{2} a more complex situation of an inhomogeneous magnetic field is realized. The spatial modulation of the magnetic field is achieved experimentally by depositing patterned gates of superconducting or ferromagnetic materials on top of heterostructures. \cite{2,3,4} An alternative approach to produce inhomogeneous magnetic fields is by varying the topography of an electron gas. \cite{5} Theoretically it is found that different functional magnetic field profiles induce diversity of magnetic edge states \cite{6,7,8,9,10,11,12,13} with remarkable time-reversal asymmetry. \cite{6,13}

Recently, studies of the combined effect of Rashba SOI and quantizing magnetic field on the carrier energy dispersions and the spin polarization properties as well as on the spin Hall effect have attracted much attention. \cite{13,12.1,14,15,16,17,18,19,20,21} The interplay between Rashba, Dresselhaus, and Zeeman interactions in a magnetic field has been also studied. \cite{22,23} However, these works are mainly confined to the uniform magnetic fields. Meanwhile, as shown in Ref. \onlinecite{26}, an inhomogeneous magnetic field can be used as an effective tool to manipulate electron spins, mainly to control spin polarized current, similar to the spin field-effect transistor.
Thus far, the efforts in this field are mainly directed towards studying the effect of SOI controlled edge transmission on the resistance of magnetic barriers. \cite{24,25,28,29} A large spin polarization $ \left(\backsim 95\% \right)$  is found in these systems with the potential to use them as spin injectors in spin-logic devices as well as magnetic sensors with ultra-high density storages.
In the present paper we investigate the magnetic edge states and their transport properties in a 2DES, exposed to a normal inhomogeneous magnetic field in the presence of SOI and the Zeeman effect. We focus on the joint Rashba SOI and Zeeman effect and study the spectral and spin transport properties of these spin interface states, their control by means of combined effect of the SOI and inhomogeneous magnetic field. Our study will be based on investigations of the single-particle spectrum of the inhomogeneous magnetic field induced edge states and their transport in 1D and 2D systems \cite{5,6,8} as well as we extend our previous study for the semiinfinite system with Rashba SOI and homogeneous magnetic field, confined by the infinite potential wall. \cite{12.1}

\begin{figure*}[t]
\centering
\subfigure[]{
\includegraphics[width=0.3\linewidth]{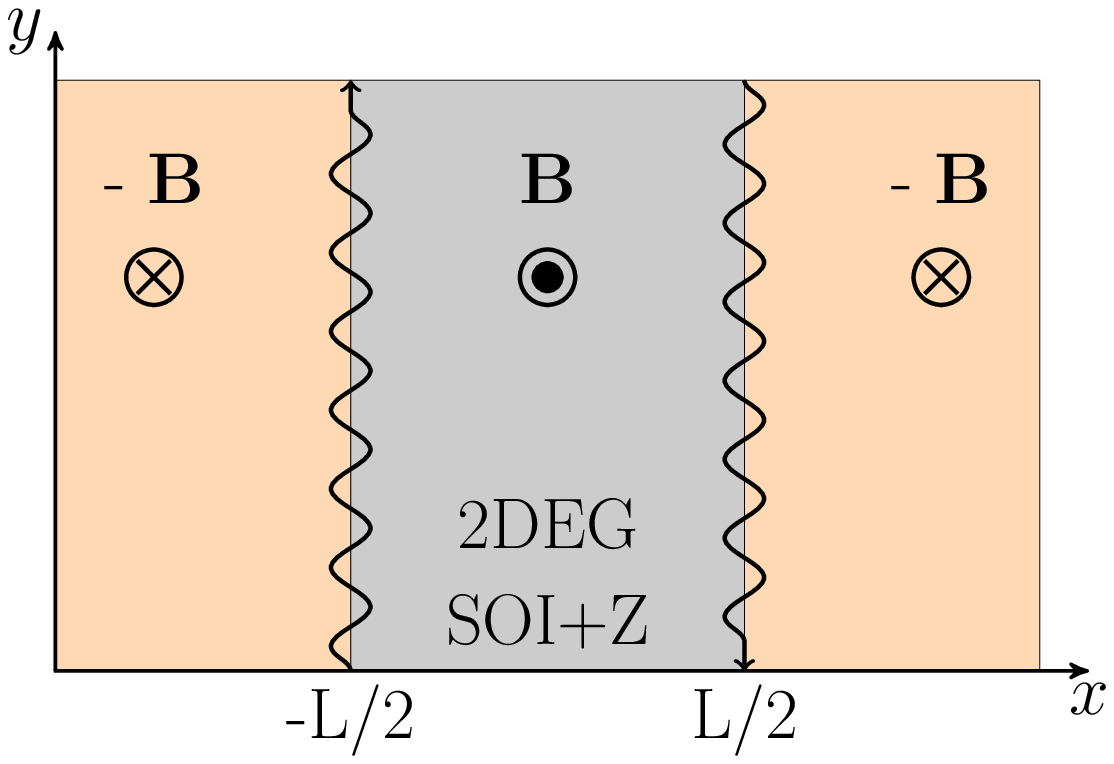}
   \label{fig.1.1}
   }
\subfigure[]{
\includegraphics[width=0.3\linewidth]{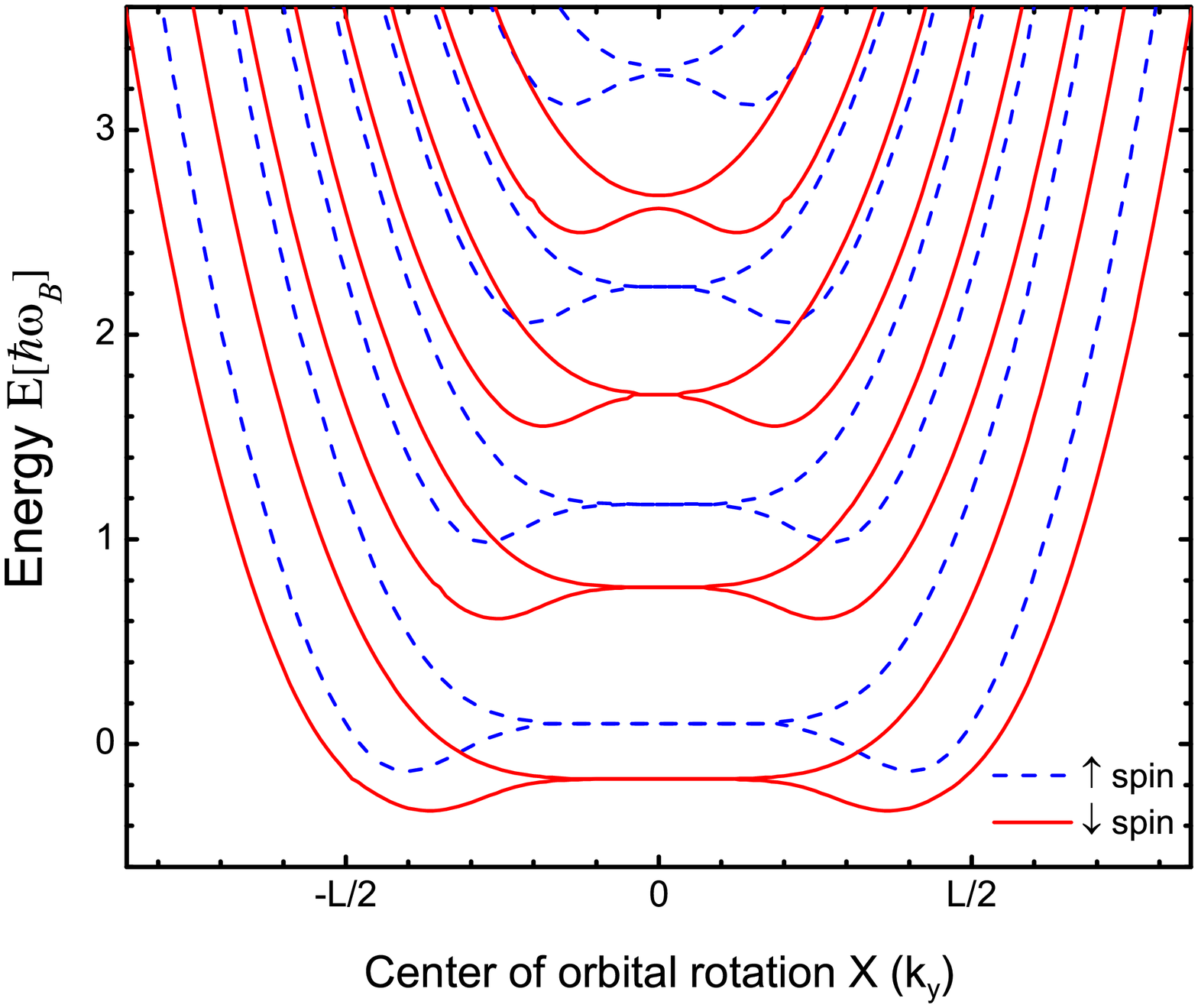}
   \label{fig.1.a}
}
\subfigure[]{
   \includegraphics[width=0.3\linewidth]{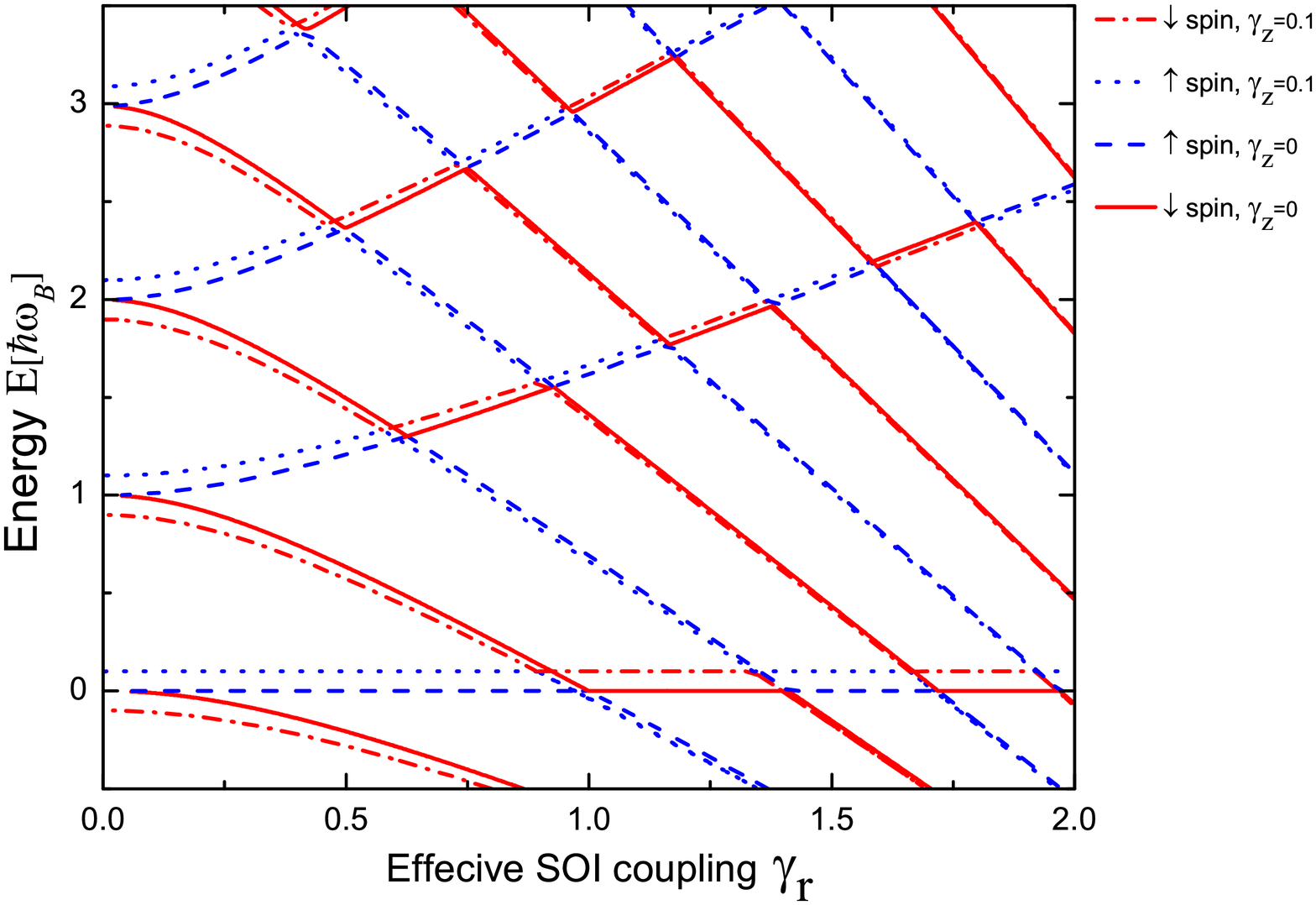}
    \label{fig.1.b}
}
\caption{(color online) (a) Schematic view of the system. (b) The energy spectrum of spin snake states as a function of the momentum $k_{y}$   induced by the Rashba SOI, Zeeman effect and the inhomogeneous magnetic field of the antisymmetric profile, $-B/B$. The $m=4$ lowest bands are shown for $\gamma_{r}=0.3,$ $\gamma_{z}=0.1.$ The solid and dashed curves correspond to the down and up spin states. (c) The energy spectrum of spin snake states as a function of the effective SOI coupling $\gamma_{r}$ for $X \left( k_{y} \right)=0.$ Dotted and dash-dotted curves correspond to spin-$\uparrow$ and spin-$\downarrow$ states for $\gamma_{z}=0.1$ and dashed and solid curves to spin-$\uparrow$ and spin-$\downarrow$ states for $\gamma_{z}=0.$}
\label{fig.1}
\end{figure*}

We study the spin edge states, induced by the combined effect of spin-orbit and Zeeman interactions and a inhomogeneous magnetic field, exposed perpendicularly to the two-dimensional electron system (2DES) in $\left( x,y\right)-$plane. We assume that the 2DES resides in a quantum well, formed in the (001) plane of a zincblende semiconductor heterostructure, and consider the case of double magnetic interfaces where the magnetic field exhibits a discontinuous jump in the transverse $x-$direction and changes its sign on the left- and the right-hand side of the magnetic interface, located at at $x=\pm L/2.$ The schematic view of the system is shown in Fig. \ref{fig.1.1}. This model can be described by an effective two-dimensional Hamiltonian of the form
\begin{equation}
H=H_{0}+H_{SOI}+H_{Z},
\label{eq.1}
\end{equation}
where the Hamiltonian of free particle in a quantizing perpendicular magnetic field is $H_{0}=\left(\vec{\Pi}^2/2m^\ast \right) \hat{\tau},$ the Bychkov-Rashba spin-orbit interaction (SOI) Hamiltonian $H_{SOI}=\alpha \left(\Pi_{x}\hat{\sigma}_{y}-\Pi_{y}\hat{\sigma}_{x}\right),$ and the Zeeman interaction $H_{Z}=\left(g\mu_B/2 \right) \vec{\sigma}\cdot\vec{B}.$ Here $m^\ast$ denotes the electron effective mass. The electron kinetic momentum operator $\vec{\Pi}=\vec{P}-\left(e/c\right)\vec{A}$ where $\vec{P}=-i\hbar\vec{\nabla}$ is the canonical momentum and the vector potential $\vec{A}$ is given by the inhomogeneous magnetic field though $\vec{B}=\vec{\nabla}\times\vec{A}=B_{z}\left(x\right)\hat{\bz}.$ The unity matrix $\hat{\tau}$ and the Pauli matrices $\hat{\sigma}_{x},$ $\hat{\sigma}_{y}$ and $\hat{\sigma}_{z}$ act in the pseudospin space. $\alpha$ is the Bychkov-Rashba spin-orbit coupling constant, $g$ is the Lande factor of electron and $\mu_{B}=e\hbar/2m_{0}c$ is the Bohr magneton where $m_{0}$ is the free electron mass. We have assumed that electrons are confined to the lowest energy
subband in the $z-$direction. In the Landau gauge so that the components of the vector potential are $\vec{A}\left(x\right)=\left(0,xB_{z}\left(x\right),0\right)$ where $B_{z}\left(x\right)=B_{0}f\left(x\right)$ with $f\left(x\right)=\text{ sgn}\left(\abs{L/2}-\abs{x}\right)$ and $\text{sgn}\left(x\right)=\left[1-2\Theta_{H}\left(x\right)\right]$ with $\Theta_{H}\left(x\right)$ denoting the Heaviside unit step function.
\begin{figure*}[t]
\centering
\subfigure[]{
   \includegraphics[width=0.3\linewidth]{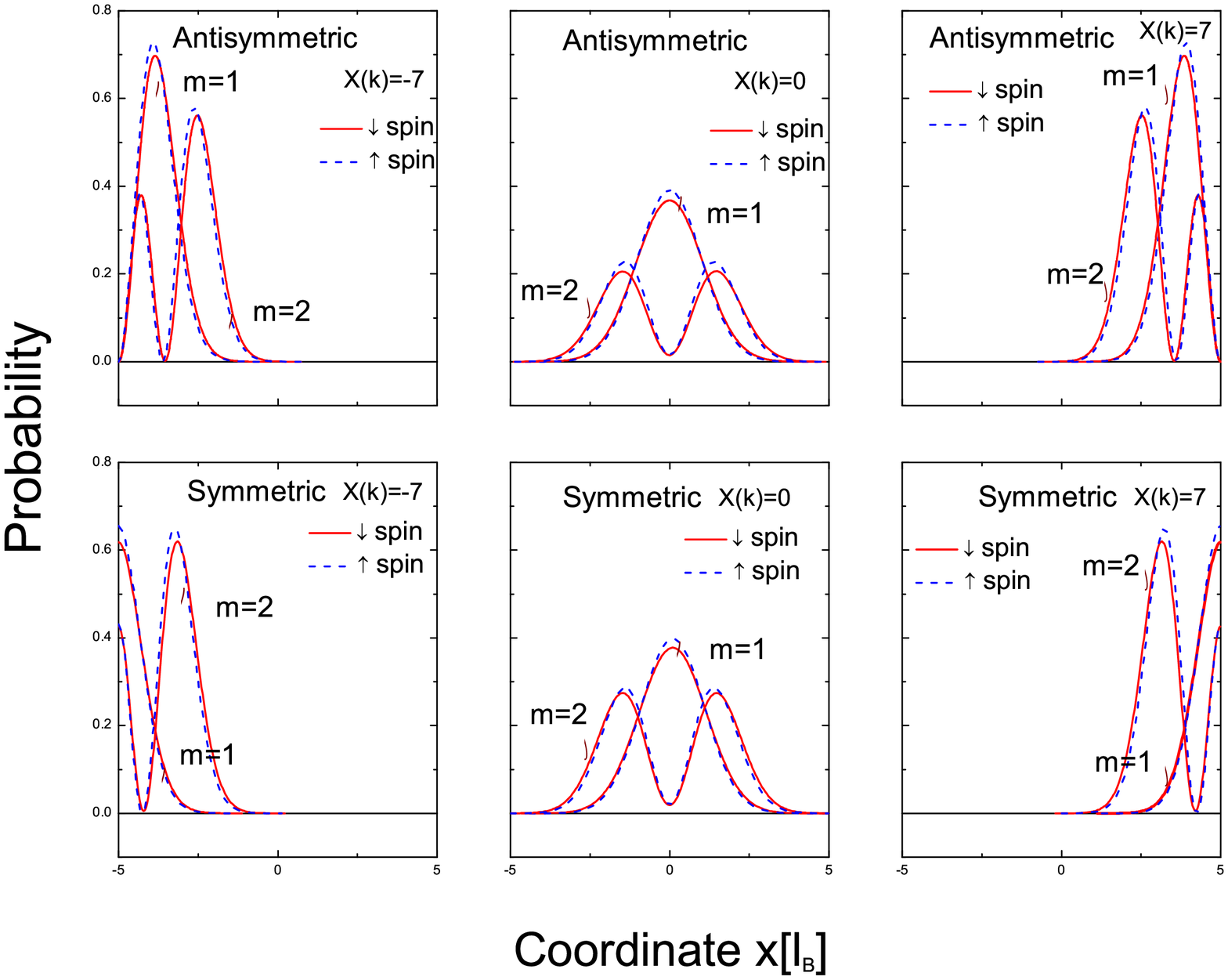}
    \label{fig.1.c}
}
\subfigure[]{
\includegraphics[width=0.3\linewidth]{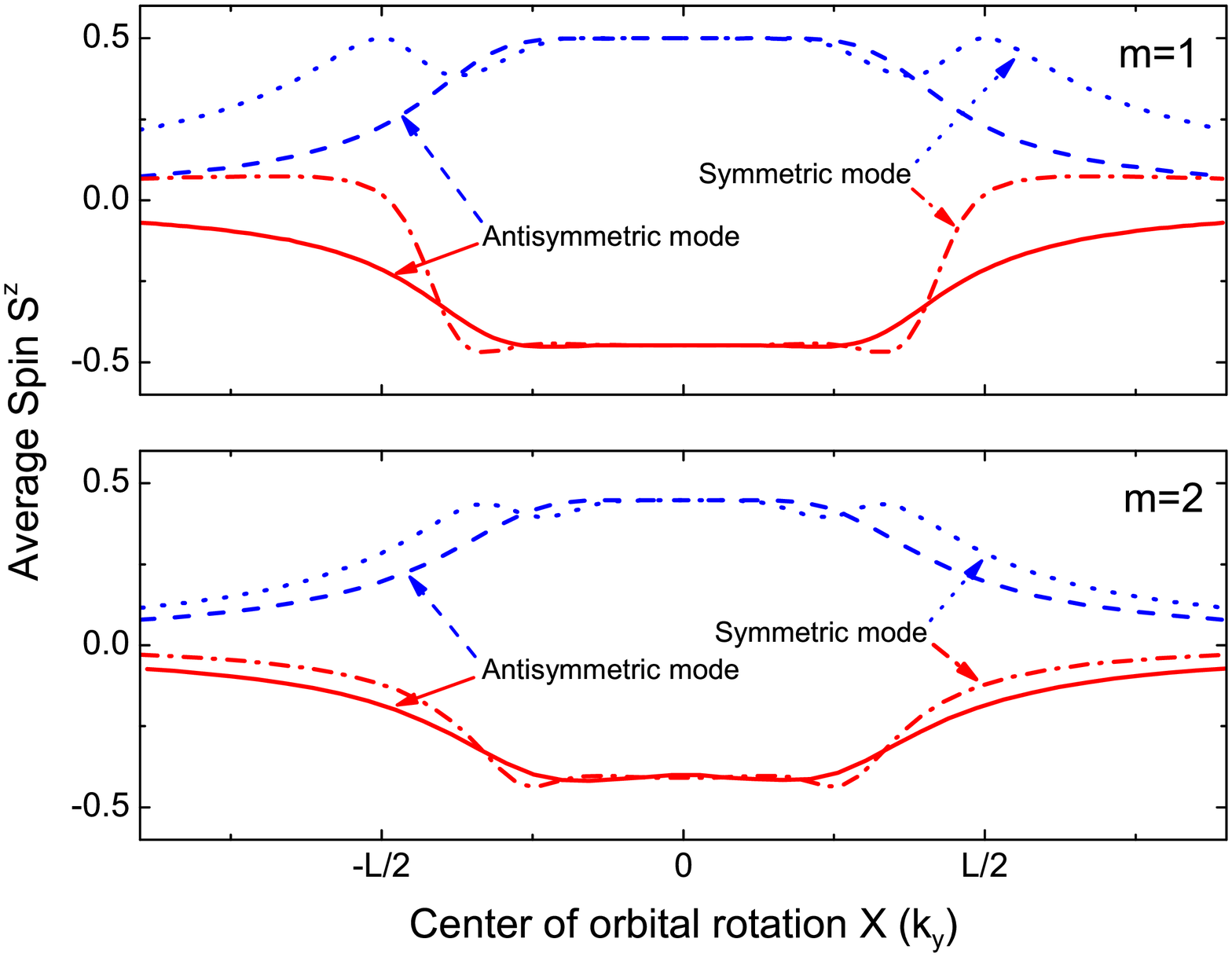}
    \label{fig.3.1}
}
\subfigure[]{
   \includegraphics[width=0.3\linewidth]{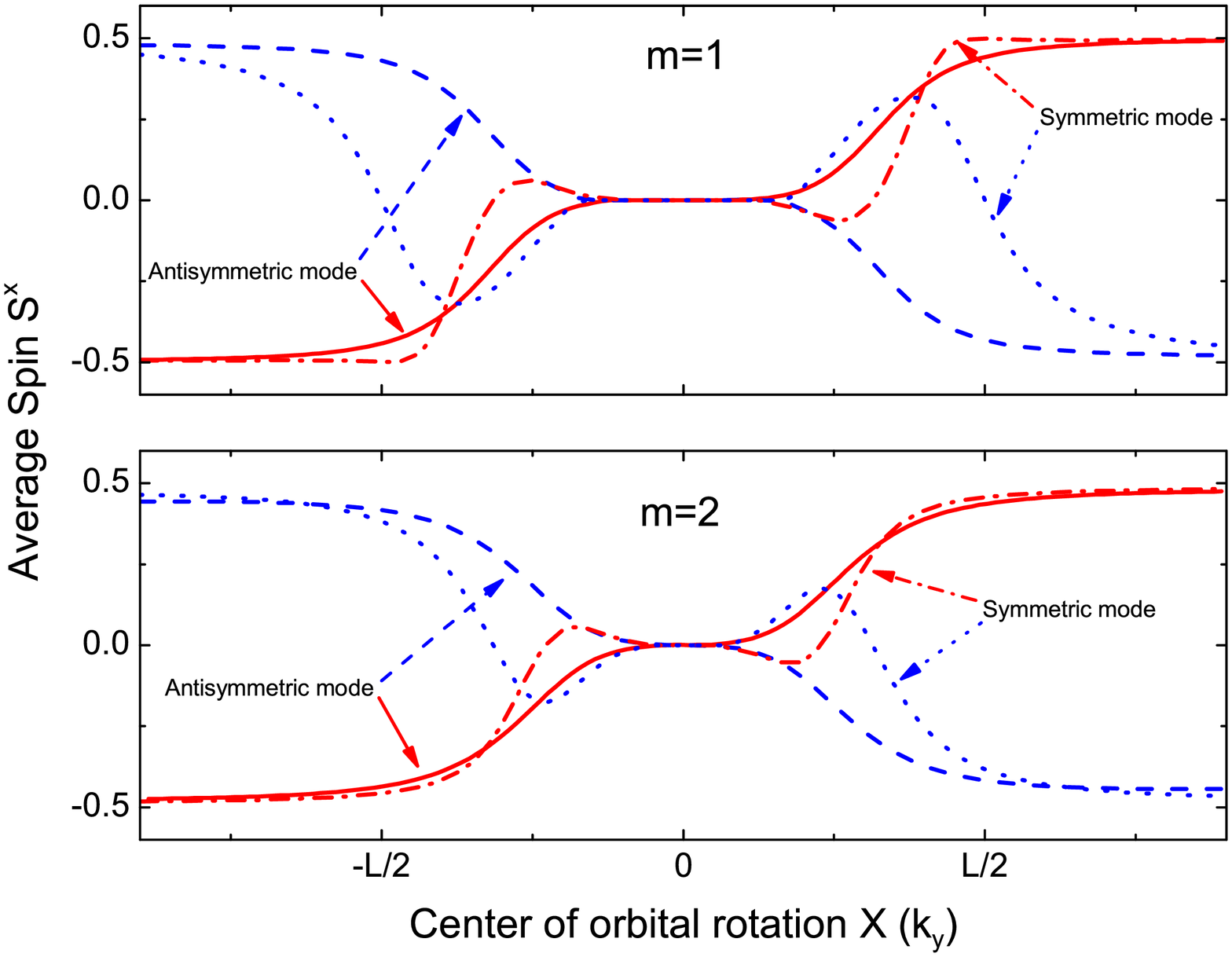}
    \label{fig.3.2}
}

\caption{(color online)  (a) The probability densities for different spins and wave vectors of the first $m=1,2$ two bands. (b) The $z-$ and (c) the $x-$components of average spins in units of $\hbar$ as a function of $X \left( k_{y} \right)$ for the first two bands $m=1,2$ and for $\gamma_{r}=0.3,$ $\gamma_{z}=0.1.$ }
\label{fig.3}
\end{figure*}
 Using the ansatz $\Psi\left(x,y\right)=e^{ik_{y}y}\chi_{k_{y}}\left(x\right),$ we can reduce two-dimensional Schrodinger equation $H\Psi=E\Psi$ to the one-dimensional problem, where $E$ is the electron total energy and $k_{y}$ the electron momentum in $y-$direction, along which the translational invariance is not broken. Taking into account explicity that the electron wave function $\chi_{k_{y}}\left(x\right)$ is spinor, $ \chi_{k_{y}}\left( \xi \right)=\begin{pmatrix} \chi_{1k_{y}}\left(\xi\right) & \chi_{2k_{y}}\left( \xi \right) \end{pmatrix}^{T}$ with $\xi=x f\left(x\right)-X\left(k_{y}\right)$ in $x-$direction should satisfy the following equation
\begin{equation}
\begin{pmatrix} h_{\nu_{-}} & \pm h_\pm \\ \pm h_\mp & h_{\nu_{+}} \end{pmatrix} \begin{pmatrix} \chi_{1k_{y}}\left( \xi\right) \\ \chi_{2k_{y}}\left( \xi \right) \end{pmatrix}=0
\label{eq.9}
\end{equation}
where the upper (lower) sign corresponds to $\abs{x}<L/2$ $\left(\abs{x}>L/2\right)$ with $B_{z}\left(x\right)=B_{0}$ $\left(B_{z}\left(x\right)=-B_{0}\right).$ In Eq. (~\ref{eq.9}) we have introduced the following operators:
\begin{align}
h_{\nu_{\pm}}=\left(\frac{d^2}{d \xi^2}+\nu+\frac{1}{2}-\frac{\xi^2}{4}\pm\gamma_{z}\right),
\label{eq.10}\\
h_\pm=-\gamma_{r} \left[ \frac{\xi}{2} \mp \frac{d}{d \xi} \right].
\label{eq.11}
\end{align}
in which the effective potential depends additionally on the wave vector
$k_{y}$ along $y-$direction. In Eq. (\ref{eq.9}) we express the energy
$E\rightarrow\left(  \nu+1/2\right)  \hbar\omega_{B}$ in units of the
cyclotron energy, $\hbar\omega_{B}\equiv\hbar eB_{0}/m^{\ast}c$, and the
length $x\rightarrow xl_{B}/\sqrt{2}$ in the magnetic length, $l_{B}%
\equiv \sqrt{\hbar c/eB_{0}}$. We introduce also the dimensionless SOI coupling
constant $\gamma_{r}=\sqrt{2}\alpha/v_{B}$ with the cyclotron velocity
$v_{B}=m^{\ast}l_{B}/\hbar,$ the Zeeman interaction constant $\gamma_{z}=gm^\ast\mu_{B}c/e\hbar$ and the dimensionless coordinate of the center of orbital rotation $X(k_{y})=\sqrt{2}k_{y}l_{B}$.

The system of equations in (\ref{eq.9}) has to be solved under the boundary conditions $\chi_{k_{y}}\left(x\right)\rightarrow 0$ when $x\rightarrow \pm \infty.$ In the absence of SOI and Zeeman effect, $\gamma_{z}=0$ and $h_\pm=0,$ we obtain the solution in terms of parabolic cylindrical functions, $D_{\nu}\left(x\right).$ In the presence of SOI we search the solution of matrix equation (\ref{eq.9}) for $\abs{x}<L/2$ as $\chi_{1k_{y}}\left( \xi \right)=a_{1} D_\mu\left( \xi \right)$ and $\chi_{2k_{y}}\left( \xi\right)=a_{2} D_{\mu-1}\left( \xi \right).$ Here $a_{1}$ and $a_{2}$ are the $x-$independent spinor coefficients to be determined by solving the system of equations obtained
from inserting this ansatz into Eq. (\ref{eq.9}). The index $\mu \geq -1/2$ in general differs from the index $\nu.$ Using the recurrent properties of the parabolic cylindrical functions \cite{12.1}
we obtain from Eq.(\ref{eq.9}) for $\abs{x}<L/2$ two (non-normalized) independent bulk solutions
\begin{equation}
\chi_{k_{y}}^\pm \left(\xi\right)= \begin{vmatrix} D_{\mu_{\pm}\left(\nu, \gamma_{r}, \gamma_{z}\right)} \left(\xi\right) \\ c_{\pm} D_{\mu_{\pm}\left(\nu, \gamma_{r}, \gamma_{z} \right)-1} \left(\xi\right) \end{vmatrix},
\label{eq.22}
\end{equation}
where $\mu_\pm \left(\nu, \gamma_{r}, \gamma_{z} \right)=\nu+\frac{1}{2}+\frac{\gamma_{r}^2}{2} \pm \sqrt{\nu \gamma_{r}^2 + \frac{1}{4} \left(1+\gamma_{r}^2 \right)^2 + \gamma_{z} \left(\gamma_{z}+1\right)}$ and $c_{\pm} \left( \nu, \gamma_{r}, \gamma_{z} \right)=-\frac{1}{\gamma_{r}} \left(\frac{1}{2}+\frac{\gamma_{r}^2}{2}\pm \sqrt{\nu \gamma_{r}^2 + \frac{1}{4} \left(1+\gamma_{r}^2 \right)^2 + \gamma_{z} \left(\gamma_{z}+1\right)} \right).$
Notice that for any value of $\nu \geq -1/2,$ the Weber equation has two independent solutions, $D_\nu \left(x\right)$ and $D_\nu \left(-x\right)$ at the same time for $\nu \neq 0,1,2,...$ Thus, there is a second set of, in general, independent solutions of Eq. (\ref{eq.9})
\begin{equation}
\eta_{k_{y}}^\pm \left(\xi\right)= \begin{vmatrix} D_{\mu_{\pm}\left(\nu, \gamma_{r}, \gamma_{z} \right)} \left(-\xi\right) \\ -c_{\pm} D_{\mu_{\pm}\left(\nu, \gamma_{r},  \gamma_{z} \right)-1} \left(-\xi\right) \end{vmatrix}.
\label{eq.25}
\end{equation}

The general solution of Eq. (\ref{eq.9}) is given by
\begin{equation}
\psi_{k_{y}}\left(\xi\right)=\alpha \chi_{k_{y}}^{+} \left(\xi\right)+\beta \chi_{k_{y}}^{-} \left(\xi\right)+\lambda \eta_{k_{y}}^{+} \left(\xi\right)+\vartheta \eta_{k_{y}}^{-} \left(\xi\right)
\label{eq.26}
\end{equation}
and choose the $\alpha,$ $\beta,$ $\lambda$ and $\vartheta$ so that the new wave function $\psi_{k_{y}}\left(\xi\right)$ and its derivative $\psi^{'}_{k_{y}}\left(\xi\right)$ be continuous at $x=\pm L/2: $ \cite{7,9} \begin{gather}
\left. \frac{d}{d \xi} \psi_{k_{y}}\left(\xi\right)\right|_{\xi=\pm L/2-X\left(k_{y}\right)}=0,
\label{eq.26.1} \\
\left. \psi_{k_{y}}\left(\xi\right) \right|_{\xi=\pm L/2-X\left(k_{y}\right)}=0
\label{eq.27}
\end{gather}
for the symmetric and antisymmetric wave function, respectively. This leads to a change in energy of the electron states, which can be understood as a lifting of the degeneracy of the two original electron wave functions.  The corresponding linear systems of equations defined by this condition have nontrivial solutions if the respective determinants vanish at $x=\pm L/2$ which gives the dispersion equations.
 The wave functions are
\begin{equation}
\psi_{k_{y}} \left( \xi \right)= \alpha \left| \begin{matrix} \psi_{1,k_{y}} \left( \xi \right) \\ \psi_{2,k_{y}} \left( \xi \right) \end{matrix} \right| ,
\label{eq.29.1}
\end{equation}
where two spinors $\psi_{1,k_{y}} \left( \xi \right)$ and $\psi_{2,k_{y}} \left( \xi \right)$ are defined from Eq. (\ref{eq.26}) with corresponding spinors of $\chi^\pm_{k_{y}} \left(\xi\right)$ and $\eta^\pm_{k_{y}} \left(\xi\right)$ from Eqs. (\ref{eq.22}) and (\ref{eq.25}), respectively and the coefficients in Eq. (\ref{eq.26}) are the eigenvectors of the (\ref{eq.26.1}) for symmetric and (\ref{eq.27}) for the antisymmetric wave functions. The normalization of the wave functions $\int{d \xi \psi^{\dag}_{k_{y}} \left( \xi \right) \psi_{k_{y}} \left( \xi \right)}=1$ gives the amplitude
\begin{equation}
\alpha= \left\{\int{ d \xi \left[ \abs{\psi_{1,k_{y}} \left( \xi \right)}^2+\abs{\psi_{2,k_{y}} \left( \xi \right)}^2 \right]}\right\}^{-1/2}.
\end{equation}
The dispersion equations from matching conditions (\ref{eq.26.1}) and (\ref{eq.27}) are quadratic with respect to the parabolic cylindric functions, therefore for a given band index, $m$, each of the equations has two solutions, $E_{sm} \left(k_{y} \right),$ corresponding to the nondegenerated magnetic edge states with the spin $s=\uparrow$ and $\downarrow .$ 
The dispersion equation arising from (\ref{eq.26.1}) gives rise to the unusual spectrum of magnetic spin edge states with the negative velocities, \cite{7,10,11} while the dispersion equation form (\ref{eq.27}) gives the spectrum of the spin edge states restricted by infinite potential wall. \cite{12.1}

Here we carry out the actual calculations of the spectrum of spin edge states, carried by snake orbits along the magnetic interfaces in a 2DES. In the presence of a perpendicular magnetic field, the efficiency of SOI is determined by the dimensionless coupling constant $\gamma_{r}$, which are inversely proportional to the square root of the magnetic field strength, $B_{0}.$ We carry the actual calculations for magnetic fields corresponding to the cyclotron splitting of about 5 K. In InAs, with 
$m^{\ast}=0.026m_{0},$ such a cyclotron splitting is achieved for $B_{0}=0.1 $T and taking the Rashba constant  $\alpha \approx 112.49  $mev\AA \cite{30} we have
 $\gamma_{r}=0.45.$ Using the value of Lande factor $g=-15$ for bulk InAs \cite{31} we calculate the Zeeman effect constant $\gamma_{z}=-0.1.$ The Rashba coupling constant can be changed by varying external electric field. 
To compare the systems with infinite hard wall \cite{12.1} and the magnetic interface induced edge states, we carry out our calculations for Rashba coupling $\gamma_{r}=0.3$ and the Zeeman coupling $\abs{\gamma_{z}}=0.1.$\cite{32} In Figs. \ref{fig.1.a} and \ref{fig.1.b} we plot the energy spectrum of spin edge states, $E_{sm}\left(k_{y}\right),$ as a function of momentum $k_{y}$ (Fig. \ref{fig.1.a}), which we obtain by solving numerically the dispersion equations for $\gamma_{r}=0.3$ and $\gamma_{z}=0.1$. As it is shown in Ref. \onlinecite{12.1}, for the edge states confined by the infinite potential wall with stronger effective SOI coupling the energy spectrum shows well-pronounced anti-crossings. In Fig. \ref{fig.1.b} we plot the symmetric mode of energy spectrum of spin snake states as a function of the effective SOI coupling $\gamma_{r}$ or, what is the same, versus $B_{0}^{-1/2}$ for $\gamma_{z}=0$ and $\gamma_{z}=0.1$ with the fixed value of the center of orbital rotation $X \left( k_{y} \right)=0.$ Dotted (dashed) and dash-dotted (solid) curves correspond to spin-$\uparrow$ and spin-$\downarrow$ states for $\gamma_{z}=0.1$ ($\gamma_{z}=0.$)
 It is seen from the figure that the anticrossings are observed also in the presence of Zeeman effect, moreover, depending on value of SOI coupling constant or magnetic field, the Zeeman effect can either increase or decrease the splitting of energy levels.
At low values of SOI coupling the splitting of energy levels is mainly due to the Zeeman effect while for the lower magnetic field SOI becomes dominant. From the energy spectrum we calculate the corresponding group velocities along $y-$direction, $v_{sm} \left(k_{y} \right)=\partial E_{sm} \left(k_{y} \right)/ \partial k_{y},$ (along $x-$direction $v_{x}=0$). As seen from the spectrum there are two types of magnetic edge states, which alternate each other. The lower located branches with the quantum numbers $n=2m$ ($m=0,1,2,...$) have local minima at some finite values of $k_{y}$ where they reverse their velocity direction while the higher located states with $n=2m+1$ show monotonic behavior in the whole variation range of $k_{y}.$ The eigenstates with a negative velocity have no classical counterpart while the eigenstates with large absolute values of $k_{y}$ represent the classical snake orbits. In the middle of the sample, these two types of branches coagulate from above and below to the dispersionless Landau levels,
labeled by the quantum number $m$ so that the eigenstates become the twofold degenerated. The inclusion of SOI splits each of these two branches with $n=2m$ and $n=2m+1$ into the spin up (dashed curves) and down (solid curves) states so that there are actually four spin snake states for $\abs{k_{y}}\rightarrow \infty$ and two spin splitted but still twofold degenerated Landau levels for $\abs{k_{y}}\rightarrow 0.$ It is seen in Fig. \ref{fig.1.a} that the spin splitting increases both with the main quantum number $m$ and with the absolute value $\abs{k_{y}}$ of the snake state wave number. 
This can result in the net spin current since at the sufficiently high Fermi energies there are more spin down than spin up current carrying states.

The probability densities for different spins and wave vectors of the first two bands are shown in Fig. \ref{fig.1.c}. The probabilities for different spins and wave vectors are different in two bands for the antisymmetric as well as for the symmetric mode. This difference gives rise to a spatial separation of the spin-$\uparrow$ and spin-$\downarrow$ states. The difference of the probabilities appears even at the center of the sample when the electrons are in the quasibulk Landau levels.
\begin{figure}
\centering
\includegraphics[width=\columnwidth]{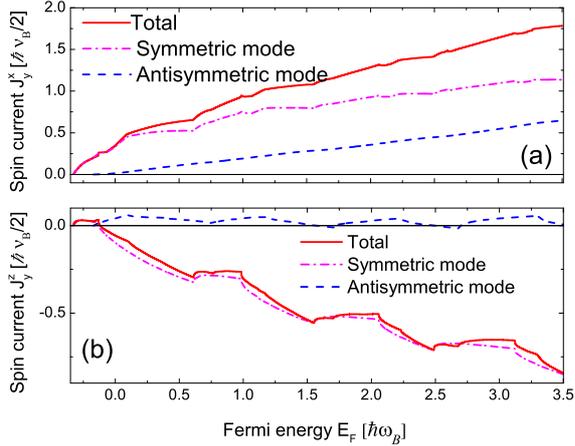}
\caption{(color online) (a) The $x-$ and (b) the $z-$component spin current for $\gamma_{r}=0.3,$ $\gamma_{z}=0.1$ as a function of the Fermi energy $E_{F}.$ The dashed and dotted curves plot the separate contributions to the antisymmetric and symmetric modes.}
\label{fig.4}
\end{figure}
Further we carry out the actual calculations of the average components of spins \begin{equation}
S^{x,z}_{sm} \left( k_{y} \right) =\left. \frac{\hbar}{2} \int^{\infty}_{0} { d x \psi^{\dag}_{k_y} \left(x\right) \hat{\sigma}_{x,z} \psi_{k_y} \left( x \right) } \right|_{E=E_{sm} \left( k_{y} \right)}
\label{eq.30}
\end{equation}
and the spin-current components, carried by the snake orbits along the magnetic interfaces of 2DES as a function of the Fermi energy
\begin{equation}
J^{x,z}_{y} \left( E_{F} \right) =\sum_{s,m}\left. S^{x,z}_{sm} \left(k_{y} \right) v_{sm} \left( k_{y} \right) \right|_{E_{sm} \left( k_{y} \right)=E_{F}}.
\label{eq.32}
\end{equation}
In Figs. \ref{fig.3.1} and \ref{fig.3.2} we plot the average spin components $S^{z}_{sm} \left( k_{y} \right)$ (Fig. \ref{fig.3.1}) and  $S^{x}_{sm} \left( k_{y} \right)$ (Fig. \ref{fig.3.2}) in units of $\hbar$ as a function of the center of the orbital rotation $X \left( k_{y} \right)$ for the first two bands $m=1,2$ when $\gamma_{r}=0.3$ and $\gamma_{z}=0.1.$ Because the transverse wave functions are real, the $y$ component of the spin vanishes identically, $S^{y}_{sm} \left( k_{y} \right)=0.$ Because in the quasibulk Landau states electrons have no preferential direction
in the $(x,y)$ plane of their cyclotron rotation \cite{12.1} at small values of $\abs{k_{y}}$ when electrons are far from the interfaces, the spins are mainly aligned along $z$ axes.
 The small splitting near the interfaces (Fig. \ref{fig.3.1}) is due to the Zeeman effect. In contrast of monotonic behavior of the average spin components for antisymmetric mode, the spin orientation of the symmetric mode shows small oscillations near the interfaces.  In the opposite limit of large $\abs{k_{y}},$ the edge channels are formed and the spins are mainly aligned in $x$ direction, perpendicular to the $y$ direction of electron propagation. It is seen that
 the electron propagation direction in $y$ axes in left and right interfaces are opposite and the SOI aligns the electrons in opposite directions in $x$ axes (Fig. \ref{fig.3.2}). One should note that due to the spin splitting the absolute values of the average spin components do not equal in the $\uparrow$ and $\downarrow$ states for antisymmetric as well as for the symmetric modes and this asymmetry becomes stronger with the band
index $m.$

In Fig. \ref{fig.4} we plot the $x$ and $z$ components of the net spin current $J^{x,z}_{y} \left( E_{F} \right)$ in $y$ direction as a function of the Fermi energy defined in Eq. (\ref{eq.32}). The dashed and dash-dotted curves plot the antisymmetric and symmetric modes, respectively. Because of the antisymmetric behavior of the $J^{z}_{y} \left( k_{y} \right)$ for any value of $E_{F}$ there are two spin current contributions: $J^{z}_{y} \left( k_{y} \right)$ and $J^{z}_{y} \left( -k_{y} \right)$ at different interfaces, which are opposite and the total spin current is zero in all range of the Fermi energy. However, in order to evaluate the $z-$spin current we calculate in Fig. \ref{fig.4} for the case of single interface in the sample. It is seen that both modes show monotonic behavior in all range of the Fermi energy. In contrast of the case of the semiinfinite system, \cite{12.1}
 there is very weak oscillation in the net spin current of the antisymmetric mode in the presence of the Zeeman effect. This is because the Zeeman effect increase the splitting between $\uparrow$ and $\downarrow$ energy states in the bulk and the group velocity increases more smoothly in the Fermi energy scale. However, there are small oscillations in the symmetric mode of $J^{x}_{y} \left( E_{F} \right).$ These small peaks appear because of the unusual behavior of the energy spectrum with the negative group velocities. As seen from Fig. \ref{fig.4} the $z-$spin current changes its sign, in addition to its peaked behavior: due to the interplay between the average spin $S^{z}_{sm} \left( k_{y} \right)$ and the velocity $v_{sm} \left(k_{y} \right).$ The peaks of the spin current have a period, determined by the cyclotron energy, renormalized in the presence of SOI and Zeeman effect. One can see from Fig. \ref{fig.4} that the main contribution to the total spin current is due to the the symmetric modes. Thus, the introduction of the magnetic interface enhances the the spin current both for the $J^{x}_{y} \left( E_{F} \right)$  and $J^{z}_{y} \left( E_{F} \right).$

In conclusion, we study  spin edge states, induced by the combined effect of Rashba spin-orbit and Zeeman interactions and inhomogeneous magnetic field, exposed perpendicularly to two-dimensional electron systems. We calculate the spectrum of the spin edge states (the spin snake orbits) in 
double-interface system and show that the magnetic spin edge states reverse their propagation direction at finite values of their momentum, what has no classical counterpart.
 Depending on magnetic field, the Zeeman effect can either increase or decrease the splitting of energy levels. In the presence of double interface the spin orintations of the electrons are opposite in $x-$direction at the left and right interfaces. The contribution of symmetric mode in total spin current is several times higher then of the antisymmetric mode, which means that the $x$ and $z$ component spin current along the interface is much stronger then in the case of infinite potential wall confinement, \cite{12.1} where only the antisymmetric mode exists.

I thank Jiang Xiao and S. M. Badalyan for fruitful discussions. This work was supported by the special funds for the Major State Basic Research Project of China (No. 2011CB925601) and the National Natural Science Foundation of China (Grants No. 11004036 and No. 91121002).

\end{document}